# EROS: Short period Cepheids in the bar of the LMC

J.P. Beaulieu

*Institut d'Astrophysique de Paris, 98 bis Boulevard Arago, 75014 PARIS, FRANCE*

**Abstract.** We present Fourier analysis at 490 nm for 87 Cepheid light curves in the bar of the Large Magellanic Cloud. The photometry has be obtained through the EROS project whose main purpose is the search for baryonic dark matter in the Galactic halo. The very high quality of the photometry and the good phase coverage enable us to obtain accurate Fourier decomposition of the ligh curves. As a consequence we confirm that the so-called s-Cepheids are overtone pulsators and represent about 30 % of our sample, whereas the Classical Cepheids are fundamental pulsators. We have identified a number of features in plots of amplitude ratios and phase differences versus period. These features are usually related to resonances between different pulsation modes. If we accept that Fourier decomposition techniques applied to light curves yield positions for the centres of resonances, then it seems that resonances occur in LMC Cepheids at the same period as in Galactic Cepheids for fundamental pulsators, whereas they appear to be at shorter periods for overtone pulsators.

## 1. Introduction

EROS (Experience de Recherche d'Objets Sombres) is a collaboration between french astronomers and particle physicists to search for baryonic dark matter in the galactic halo in the form of compact objects through microlensing effects (Paczyǹski, 1986) on stars in the Large Magellanic Cloud (LMC).

A compact object in the Galactic halo, passing close enough from the line of sight to a background star in the LMC, induces an increase in the apparent brightness of the star. This phenomenon occurs owing to the alignment of the observer, the deflector and the background star. It is achromatic, symmetric in time and transient. Assuming a standard corotating halo, the time scale, $\tau_0$, of an event is given by the relation $\tau_0 = 70 \sqrt{\frac{M}{M_\odot}}$ days, where $M$ is the mass of the deflector. Therefore in order to be sensitive to a wide range of mass for compact objects, two complementary approaches have been developed.

The first involves photographic monitoring of 6.4 million stars over a 5 degrees x 5 degrees area using the ESO Schmidt telescope. Plates are taken no more than nightly in two colours, $B_J$ and $R_C$. Isolated stars can be found down to magnitude 20 with a typical photometric uncertainty of 0.3 magnitude. For the period 1990-1994, about 380 plates in blue or red have been taken and the light curves for all the objects have been built.

The second approach uses a 0.4 m f/10 reflecting telescope and a mosaic of 16 buttable CCDs covering an area of 1 degree x 0.4 degree centered in the bar of the LMC. Between December 1991 and April 1994 about 15 000 images were taken in two broad bandpass filters $B_E$ and $R_E$ centered respectively on 490 and 670 nm. We have obtained 120 000 light curves, with as many as 48 points in a given night.

General descriptions of the observing program and the CCD camera can be found in Aubourg et al. (1993a, 1993b) and Arnaud et al. (1994a, 1994b).

It is clear that EROS, like MACHO (see Cook in this volume), is specially designed to search for the fingerprint of baryonic dark matter through microlensing effects. Therefore, the study of variable stars in the LMC is not the primary goal, but a by-product. On the other hand, the extremely large amount of data collected is of great interest for systematic studies of variable stars and stellar populations in this nearby galaxy.

## 2. Search for variable stars

### 2.1. Erratic photometric fluctuations

Various software filters have been designed to search for non-periodic fluctuations like microlensing events. They are extremely powerful for the detection of stars with bursts of brightness or irregular variations. To summarize, they search for 'bumps' in a window of given duration in the stars' photometric time sequences.

### 2.2. Periodic variations

A special effort has been made to search for variable stars in a systematic and automatic way. In order to be able to analyze as much information as possible, a new algorithm has been devised which is able to detect periodicities in light curves of arbitrary shape by use of orthogonal combinations of Fourier harmonics. This technique, developed by Grison (1994), is in fact a generalization of periodogram analysis (Scargle 1982). For a group of test periods we are searching for the best parameters for a periodic-function model $\sum_{i=1}^{M} A_i \cos(i\omega t) + B_i \sin(i\omega t)$, where the coefficients are determined through the mean square method. It allows calculation of a function representative of the measurements' dispersion relative to the model, and therefore yields a confidence level for each trial period $\omega$. For each selected star the following parameters are kept in our data base: a probability of false detection, the value of the most probable period, and the first coefficients of the Fourier series of the best periodic function fitted to the measurements.

The test results for the Schmidt plates and CCDs are as follows.

SCHMIDT PLATES: The variability limit is comparable with the photometric precision, and periodicities between 0.5 days and one year are reachable. About 10 000 variable stars are expected with a confidence level better than 99% that the variations are not due to measurement fluctuations.

CCD DATA: With the 1991-1992 data we are able to detect amplitude variations of 0.3 times the photometric precision with periodicities as low as one hour. About 2000 variable stars are expected with a confidence level better than 99 %. With the 1991-1994 data, we will be able to detect amplitude variations down to 0.1 times the photometric precision.

## 3. Search for Cepheids in the bar of the LMC

### 3.1. Presentation of the 1991-1992 CCD data set

We have obtained $\sim$2000 images spanning $\sim$100 days in 1991-1992. Roughly equal numbers were obtained in $B_E$ and $R_E$. Photometric reductions were performed with a specially-written program of reconstruction and photometry. We built a catalogue comprising 80 000 stars. Then, we searched for periodic photometric fluctuations using the Grison's periodogram method. We detected about 1000 variable stars, of which 79 are eclipsing binaries (Grison et al., 1995), 102 are Cepheids ( Beaulieu et al., 1995), about 20 are RR Lyrae stars, and hundreds are red giant variables. In the next sections we will summarize our analysis of the Cepheids we have detected in the bar of the LMC (Beaulieu et al., 1995).

### 3.2. Cepheid selection

We have plotted a colour-magnitude diagram in which we delineated the limits of the Cepheid instability strip via visual inspection of a sample of light curves. We considered only stars with $B_E < 17.6$ in order to reject RR Lyrae stars. Then we performed a visual inspection of all the light curves in order to avoid eclipsing binaries. The resulting 102 stars form our sample of LMC-bar Cepheids. Only 29 of them were previously recorded in the Harvard catalogue by Payne-Gaposchkin (1971).

Some of the objects present clearly anomalous light-curve shapes, or markedly larger scatter than other stars of similar period. In these objects we searched for the most probable period, fit a Fourier model, subtract the Fourier model to the data and search for a second periodicity. Only one star (EROS2040, HV970) showed a second period. The ratio between the two period is $0.710 \pm 0.001$, which leads us to conclude that this star pulsates in both its fundamental and first-overtone mode. This star has been also classified as 'beat Cepheid,' with the same periods, by Welch using data from the MACHO project (Alcock et al., 1995, Welch in this volume).

In the following, we will concentrate on the 'non-anomalous' stars of our sample with periods less than 10 days.

## 4. Fourier analysis of Cepheid light curves.

In the last decade, many authors have shown that coefficients derived from Fourier decomposition of light curves give a quantitative basis on which to compare observations with non-linear hydrodynamic models of variable stars and results from the amplitude-equation formalism. Fourier coefficients are a powerful tool for mode discrimination and to study the presence and effect of resonances in pulsating stars (e.g. Simon and Lee (1981), the review by Simon (1988) and reference therein, Poretti (1994) and references therein, Moskalik et al. (1992), Buchler and Moskalik (1994)).

Like earlier authors, we adopt a Fourier decomposition of form

$$X_0 + \sum_{i=1}^{M} X_i \, \cos(i\omega t + \Phi_i),$$

and define the amplitude ratio

$$R_{k1} = \frac{X_k}{X_1} \quad k > 1$$

and the phase difference

$$\Phi_{k1} = \Phi_k - k \, \Phi_1 \quad k > 1.$$

As shown by Stellingwerf and Donohoe (1986), the amplitude ratio reflects the asymmetry of the variation. In other words, the larger the $R_{k1}$, the greater the departure from a sinusoidal form of the light curve. However, as was shown by earlier authors, the $R_{k1}$ can also be strongly affected by the presence of resonances between the pulsational normal modes.

The phase difference $\Phi_{k1}$ has a qualitative meaning in terms of the width of the light curve at half maximum light (Stellingwerf and Donohoe 1987). A smaller value of $\Phi_{k1}$ implies a more acute light curve. Like $R_{k1}$, $\Phi_{k1}$ can also be strongly affected by the presence of periods resonances among the normal modes of stars.

A type of low amplitude, almost sinusoidal Cepheids are classified in the General Catalogue of Variable Stars (Kholopov 1985) as s-Cepheids. It is suggested that they are fundamental mode pulsators during the first crossing of the instability strip, or first overtone pulsators.

While various terminology have been employed in the past, we shall adopt this nomenclature. Simon and Lee (1981) were the first to make a mapping of the Hertzprung progression of the changing form of classical Cepheids light curves through the use of Fourier decomposition. Then, Antonello and Porretti (1986) showed that the Galactic s-Cepheids follow different sequences in the $R_{21} - P$ and $\Phi_{21} - P$ plane. They propose a separation between classical and s-Cepheids based on the $\Phi_{21} - P$ plane, and interpretated the differences as a consequence of different pulsation modes : fundamental radial mode for classical Cepheids, and first overtone radial mode for s-Cepheids. On the other hand Gieren et al. (1990) propose that at least some of the s-Cepheids are fundamental mode pulsators.

In Figure 1, the distinction between the skew, higher amplitude and possibly bumpy light curves of classical Cepheids and the more sinusoidal, lower amplitude ones of s-Cepheids can be seen.

### 4.1. Classification

Mindful earlier works, we used the $R_{21} - P$ plane to discriminate between Classical and s-Cepheids (Figure 2). Let us recall that a low value of $R_{21}$ indicates a more-sinusoidal light curve. Therefore, we defined as s-Cepheids the 30 stars that lie under $R_{21} < 0.3$ for $P < 3$ d and $R_{21} < 0.2$ for $3 < P < 5.5$ d.

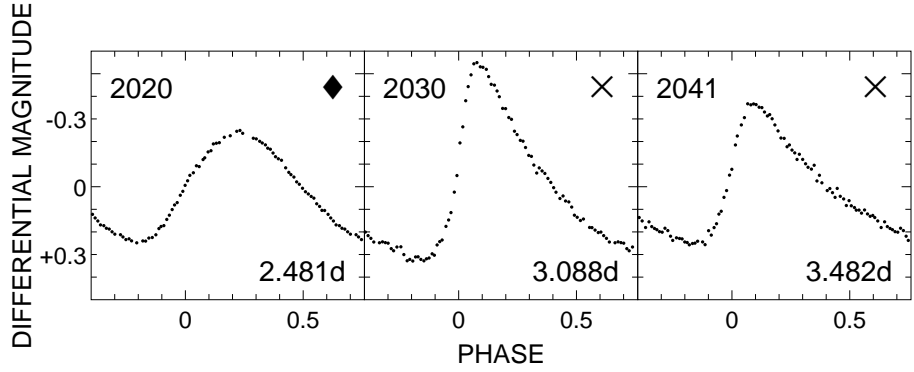

Figure 1. $B_E$ mean light curves for three Cepheids of our sample. The following symbols are used: Classical light curves (diagonal crosses), s-Cepheids (filled diamonds).

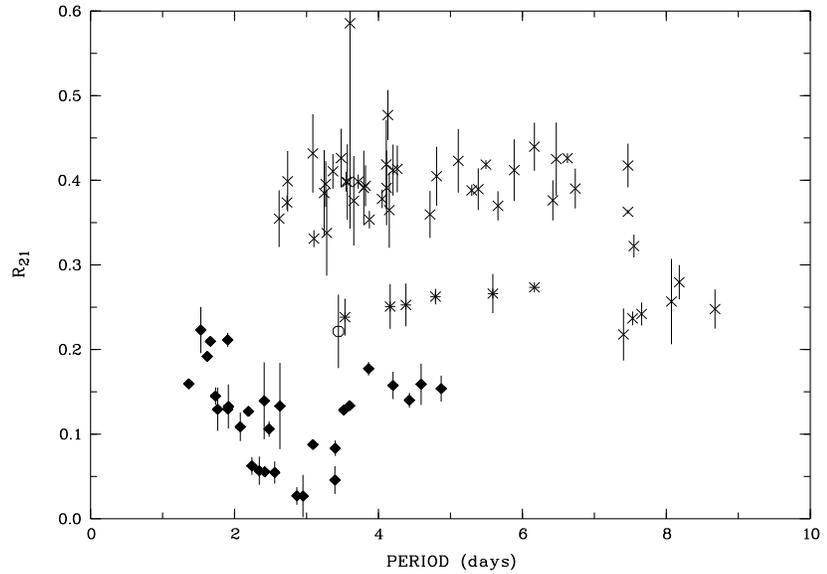

Figure 2. The $R_{21} - P$ diagram for $B_E$ light curves of the 87 objects with $P < 10$ d. The following symbols are used: Classical light curves (diagonal crosses), s-Cepheids (filled diamonds), double-mode Cepheid (open circle) and intermediate objects in the $R_{21} - P$ plane (asterisks).

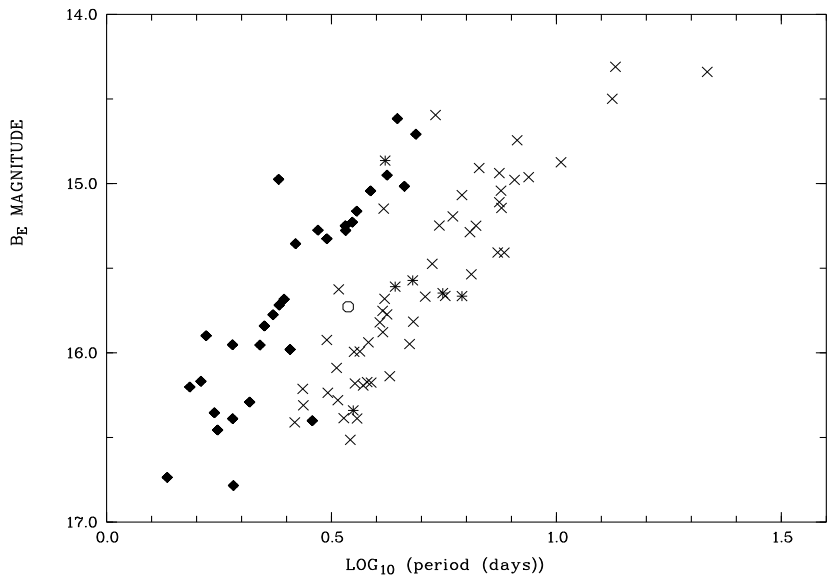

**Figure 3.**  Period-luminosity relation for Cepheids in the $B_E$ filter. The same symbols are used as in Figure 2.

A group a six stars with $R_{21} \sim 0.25$ and $4 < P < 6.5$.d. stand apart as of uncertain appartenance in $R_{21} - P$ plane. We defined them as 'intermediate stars'.

The remaining 51 stars are classified as Classical Cepheids.

We find that our classification based on morphological criteria (Fourier component values) is mirrored by a clear separation in two sequences when we plot the period-luminosity relation (Figure 3). We note that at a given period the s-Cepheids are $\sim 1$ mag. brighter. An obvious interpretation is that the upper sequence correspond to first-overtone pulsators, and most of our s-Cepheids pulsate in this mode, while the lower sequence corresponds to fundamental-mode pulsators. Most of our Classical Cepheids are pulsating in this mode.

A few stars deviate from the sequence to which they may be supposed to belong according to the classification based on Figure 2. For most of them, the images of the stars are oval, a signature of optical binarity. For some overtone pulsators, another alternative is that some of them are pulsating in the second overtone. However, it is not possible to give a conclusive answer from the available data.

The stars we defined as intermediates stars are a mixed bag. Five of them are fundamental-mode pulsators, one is an overtone pulsator, and one is a beat Cepheid.

### 4.2.  Period-luminosity relation

We plotted a period-luminosity diagram where the periods of the overtone pulsators are converted to expected fundamental-mode values through the classical relation $\frac{P(1H)}{P(1F)} \sim 0.71$. The two sequences are now closer, but not coincident. We applied a classical colour correction as defined by Feast (1984, and in this volume) to all the stars. This 'modified-period'-luminosity-colour relation is plotted in Figure 4. All the stars obey the same period-luminosity-colour relation. This confirms that **the s-Cepheids are overtone pulsators** and that **the same colour correction applies to overtone pulsators and fundamental-mode pulsators**.

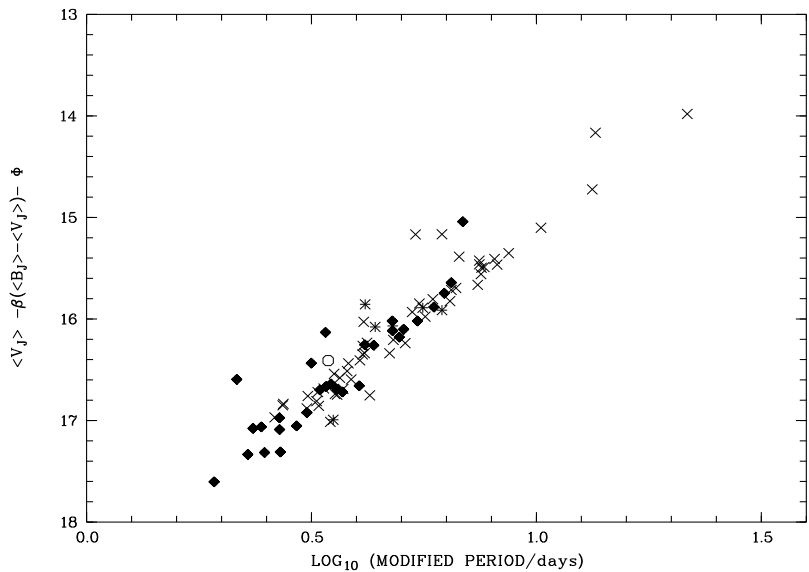

Figure 4. Period-luminosity-colour relation. The periods were modified as described in the text. The ordinate is the numerical value $<B_E> -2.95(<B_E> - <R_E>) + 2.39$. The same symbols are used as in Figures 2 and 3.

### 4.3. $\Phi_{21} - P$ plane for classical Cepheids

The classification defined in $R_{21} - P$, and mirrored in period-luminosity relation, is also observed in the $\Phi_{21} - P$ plane plotted in Figure 5. The classical Cepheids lie along a sequence corresponding to the so-called Hertzprung progression.

The extremely regular progressions in $R_{21} - P$, $\Phi_{21} - P$ and other Fourier components diagrams for classical Cepheids is now well understood. The break around 10 days, observed for Galactic (Simon and Lee 1981) and Magellanic Cepheids (Andreasen and Petersen 1987) is due to the chance 2:1 period resonance between the fundamental and the second overtone in the normal-mode period spectrum of these stars. It has been understood through the formalism of amplitude equations (Buchler and Goupil, 1984, Klap et al. 1985) and reproduced by the non-linear calculations of Moskalik et al. (1992) with the new OPAL opacities.

With the present data we can only infer that the resonance lies between 8.7 and 10.2 days, as pointed out by earlier authors. A better position will be obtained via analysis of a bigger sample.

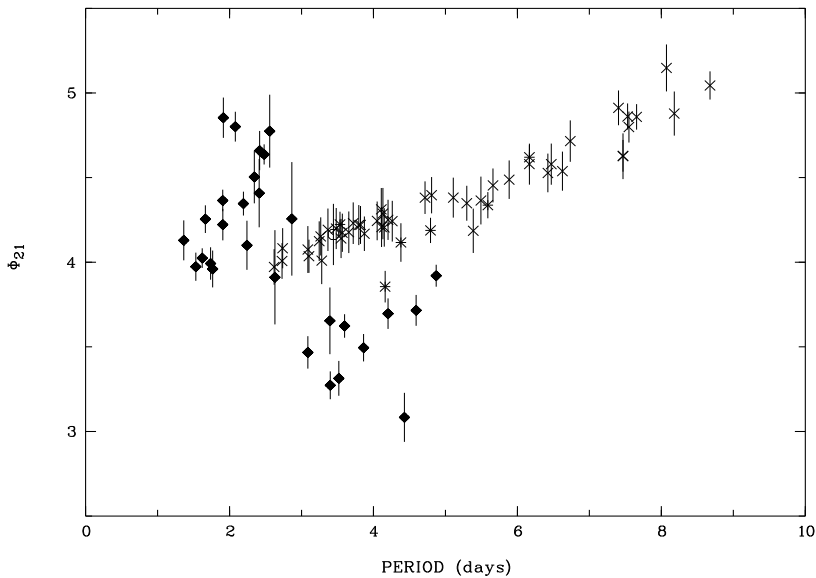

Figure 5. The $\Phi_{21} - P$ diagram for $B_E$ light curves for the 87 objects with $P < 10$ d. The same symbols are used as in Figures 2-4.

### 4.4. $\Phi_{21} - P$ plane for overtone Cepheids

The overtone pulsators fall in two well defined sequences, as was noted by Antonello et al. (1990) for Galactic Cepheids. The upper sequence is characterized by $P < 3.2$ d and $\Phi_{21} > 4.2$ rad, while for the lower sequence $P > 3.2$ d and $\Phi_{21} < 4.0$ rad. With a little calligraphic licence we called these two sequences and their possible link, the 'Z-shape'. We confirmed that all the stars that belong to the Z are overtone pulsators. Therefore, the alternative hypothesis of Gieren et al. (1990) that the stars of the lower part of the Z are fundamental pulsators is refuted.

Mindful of the effect that the 2:1 resonance between the fundamental mode and the second overtone for Classical Cepheids around 10 days has on Fourier components, the Z-shape has been interpreted by earlier authors (e.g. Poretti 1994 and references therein) as the signature of a resonance between the first and the fourth overtone.

From our sample we estimate that the Z-shape is centered around $2.7 \pm 0.2$ d, whereas it is observed around 3.2 d in the Galaxy by Porreti (1994). Buchler and Moskalik (1994) observed that in the SMC it should occur at lower period than in the Galaxy, but data quality could not precisely yield the resonance center period. A deeper comparison between the Z-shape obtained for these 3 galaxies would be of great interest for observational study of metallicity effects on this probable resonance, followed by a modelling attempt using non-linear calculations.

### 4.5. Signature of resonances for classical Cepheids around 7.5 days

For the classical Cepheids, a sharp dip feature is observed in the $R_{21} - P$ plane at $\sim 7.5 \pm 0.2$ d. It is weakly visible in $\Phi_{21} - P$, and more and more visible when going to higher-harmonic Fourier components. We think that it may be related to the 3:1 resonance between the fundamental and fourth overtone, like the feature found by Moskalik et al. (1992) in their non-linear calculations around 7.7 d and related to a dip in $\Phi_{41} - P$ found by Simon (1985), and also discussed by Antonello (1994).

## 5. conclusion

We have Fourier analyzed Cepheid light curves in the bar of the LMC using the high quality CCD photometry obtained by the EROS collaboration. We selected 87 Cepheids variables with period less than 10 days, from which 26 of them were recognized in the Harvard catalogue. We adopted a classification based on morphological criteria using Fourier components and defined as s-Cepheids the stars with low $R_{21}$ value and period less than 5.5 days. One star (EROS 2040, HV 970) is a double mode beating in the fundamental mode and the first overtone mode. 6 stars present $R_{21} \sim 0.26$ and are classified as "intermediate Cepheid". The rest are classified as classical Cepheids.

The s-Cepheids and the classical Cepheids follow different loci while ploting the period luminosity relation, and that lead us to conclude that the s-Cepheids are overtone pulsators (about 30 % of the sample ) whereas the classical Cepheids are fundamental pulsators.

The features exhibited in the $R_{21} - P$ and $\Phi_{21} - P$ plane are similar to those of galactic Cepheids. The 2:1 between the fundamental mode and the second overtone for classical Cepheids occur between 8.7 anf 10 days as it was already pointed out by earlier authors. A sharp dip feature is observed for classical Cepheids in $R_{21} - P$ around 7.5 days and we think it can be related to the 3:1 resonance between the fundamental and the fourth overtone for classical Cepheids. (This is the first detection of this resonance in the LMC, and it is found at nearly the same period as in the Galaxy.)

The Z-shape for overtone pulsators near 3 days, interpretated as the signature of the 2:1 resonance between the fourth and the first overtone is clearly seen, but is observed at a lower period ($2.7 \pm 0.2d$) than for the galaxy (3.2 d). This may be due to the lower metallicity of the LMC.

We thus find that the posistion of the resonance in overtone pulsators is shifted in the LMC, whereas for fundamental pulsators the values are similar to galactic ones.

**Acknowledgments.** This work has been done with P. Grison, J. Pritchard and W. Tobin. I thank J. Percy for inviting me to the UAI 155 colloquium. I would like to thank M. Feast, R. Ferlet, M.J. Goupil, D. Sasselov and A. Vidal-Madjar for useful discussions.